\documentclass[a4paper,amsfonts,amssymb,amsmath,reprint,showkeys,nofootinbib,twoside,superscriptaddress]{revtex4-1}
\usepackage[english]{babel}
\usepackage[utf8]{inputenc}
\usepackage{subfigure}
\usepackage[colorinlistoftodos, color=green!40, prependcaption]{todonotes}
\usepackage{amsthm}
\usepackage{mathtools}
\usepackage{physics}
\usepackage{xcolor}
\usepackage{graphicx}
\usepackage[left=23mm,right=13mm,top=35mm,columnsep=15pt]{geometry} 
\usepackage{adjustbox}
\usepackage{placeins}
\usepackage[T1]{fontenc}
\usepackage{lipsum}
\usepackage{csquotes}

\usepackage[pdftex, pdftitle={Article}, pdfauthor={Author}]{hyperref} 
\begin{document}
\title{Transverse Distance Estimation with Higher-Order Hermite-Gauss modes}

\author{Dilip Paneru}
\email[Corresponding author: ]{dpane014@uottawa.ca, dilippaneru@gmail.com}
\affiliation{Nexus for Quantum Technologies, University of Ottawa, Ottawa ON, Canada, K1N 5N6}

\author{Alessio D'Errico}
\affiliation{Nexus for Quantum Technologies, University of Ottawa, Ottawa ON, Canada, K1N 5N6}
\affiliation{National Research Council of Canada, 100 Sussex Drive, Ottawa ON, Canada, K1A 0R6}

\author{Ebrahim Karimi}
\affiliation{Nexus for Quantum Technologies, University of Ottawa, Ottawa ON, Canada, K1N 5N6}
\affiliation{National Research Council of Canada, 100 Sussex Drive, Ottawa ON, Canada, K1A 0R6}
\affiliation{Institute for Quantum Studies, Chapman University, Orange, California 92866, USA}

\date{\today} 

\begin{abstract}
We explore the use of higher-order Hermite-Gauss modes for sensing optically induced transverse displacements. In the small-displacement regime, we show that projective measurements onto the two neighboring spatial modes yield optimal Fisher information, linearly scaling with the mode order $m$. We further extend the analysis to arbitrary displacement values and derive general expressions for the Fisher information, demonstrating that higher-order modes continue to outperform the fundamental Gaussian mode even at larger separations. This approach enables enhanced displacement sensitivity with only a minimal number of measurements, offering a simple and scalable alternative to conventional Spatial Mode Demultiplexing schemes. We provide a proof-of-principle experimental demonstration using spatial light modulators, showing an order-of-magnitude reduction in estimation variance when employing Hermite-Gauss modes of order $m = 8$ and $m = 17$. These results highlight the potential of structured light for ultrasensitive displacement sensing and may enable new applications in birefringence measurements with broadband or low-coherence light sources.
\end{abstract}

\keywords{Hermite-Gauss, Superresolution, SPADE, parameter estimation}

\maketitle
Estimating transverse displacements is an essential task in numerous technological applications. In traditional imaging systems, resolution is linked to the precision with which one can determine the separation between two-point sources. This precision, in turn, is fundamentally limited by diffraction effects governed by the wavelength of light and the aperture size of the optical system~\cite{LordRayleigh}. Conventional intensity-based measurements suffer from vanishing Fisher information as the displacement between sources approaches zero, effectively linking Rayleigh’s curse to an intrinsic loss of information. The recognition of this resolution limit has driven research into alternative imaging techniques that surpass the optical Rayleigh limit, including near-field imaging~\cite{Scan,NearField} and electron-based methods, such as scanning and transmission electron microscopy~\cite{SEM, STM}. There is also growing interest in monitoring small transverse displacements. Approaches based either on incoherent methods, such as the Moiré effect \cite{kafri1990physics}, or coherent ones, such as linear photonic gears \cite{barboza2022ultra,zang2022ultrasensitive,zang2024high} or super-oscillations \cite{yuan2019detecting,Andrew:25}, have been demonstrated.  An alternative measurement scheme based on spatial mode demultiplexing (SPADE) was first proposed in ref.~\cite{OGTsang}.  In contrast with direct imaging, SPADE utilizes projection on spatial modes tailored to the Point Spread Function (PSF) of the optical system, enabling the extraction of maximal Fisher information regardless of separation distance. This conceptual breakthrough has since inspired extensive theoretical and experimental developments, leading to adaptations of SPADE for various imaging systems~\cite{ArbitraryPSF,ArbitraryPSF_Luis,French,Steinberg}, coherent sources~\cite{Coherent1,Coherent2,Coherent3}, and studies on noise-related limitations~\cite{FrenchCrossTalk,Noise}. Furthermore, SPADE has been extended to multi-parameter and higher-dimensional estimation problems~\cite{MultiParameter,2DFI}, with 2D imaging simulations~\cite{Kaden} and experimental reconstructions--leveraging techniques such as deconvolution and machine learning~\cite{Oxford} --demonstrating resolution improvements beyond the Rayleigh limit. These findings reinforce the connection between two-point source distinguishability and overall 2D imaging resolution.

\begin{figure}[t!]
  \centering 
  \includegraphics[width=\columnwidth]{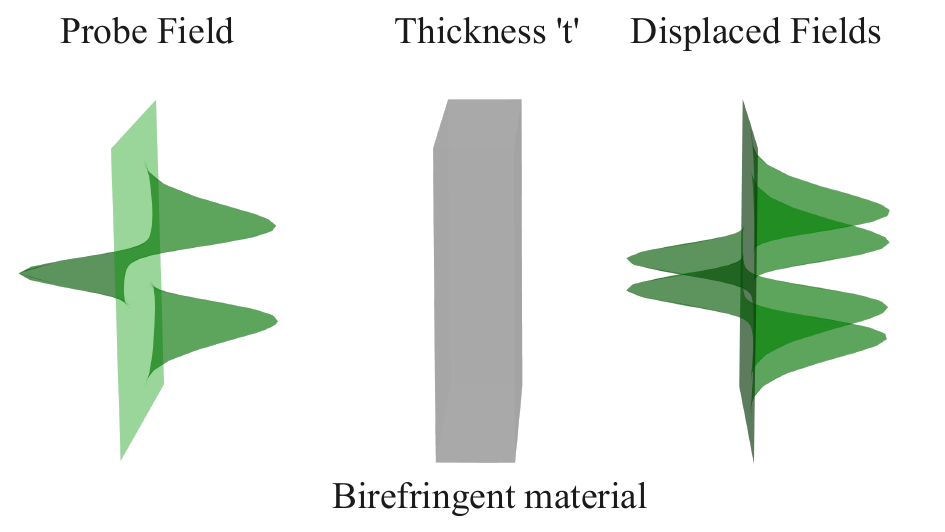}
  \caption{\textbf{Conceptual Illustration.} A schematic demonstrating how higher-order Hermite–Gauss modes can enhance sensitivity in transverse displacement estimation. Unpolarized light, prepared in a specific spatial mode (here, Hermite–Gauss mode with order $m = 2$, $n = 0$), is incident on a birefringent material. Due to the difference in refractive indices along the fast and slow axes, the beam experiences a transverse shift proportional to the material’s thickness $t$. As shown in the manuscript, using a higher-order input mode and projecting onto the two adjacent Hermite–Gauss modes yields a sensitivity enhancement that scales with the mode order.}
  \label{Fig1}
\end{figure}

Here, we examine the problem of transverse distance estimation within the framework shown in Fig.~\ref{Fig1}, considering a probe field whose spatial profile can be tailored as desired. Specifically, we aim to detect the transverse displacement caused by anomalous refraction through a birefringent material, where the displacement is directly proportional to the material’s thickness $t$. We theoretically calculate the Fisher information associated with this separation when the probe fields are the Higher-Order Hermite-Gauss modes. We find that the optimal mode projections for finding the separation consist of two nearest-neighboring modes of the input probe mode. Fisher information is evaluated in both the small- and large-separation regimes. In the small-separation regime, we find that the Fisher information scales linearly with the input mode number $m$, as expressed in Eq.~\ref{FIm}, thereby showing that higher-order modes are significantly sensitive to distance estimation. We also address the situation where one is experimentally limited by the numerical aperture of the optics involved, and show that even in such cases, the factor of $\sqrt{m}$ enhancement persists.  These findings suggest a promising method toward ultrasensitive parameter estimation using structured light. We further support our theoretical analysis with a proof-of-concept experiment, which demonstrates an order-of-magnitude improvement in sensitivity compared to measurements using the fundamental mode.

\begin{figure}[t!]
  \centering
  \hspace*{-0.5cm} 
  \includegraphics[scale=0.55]{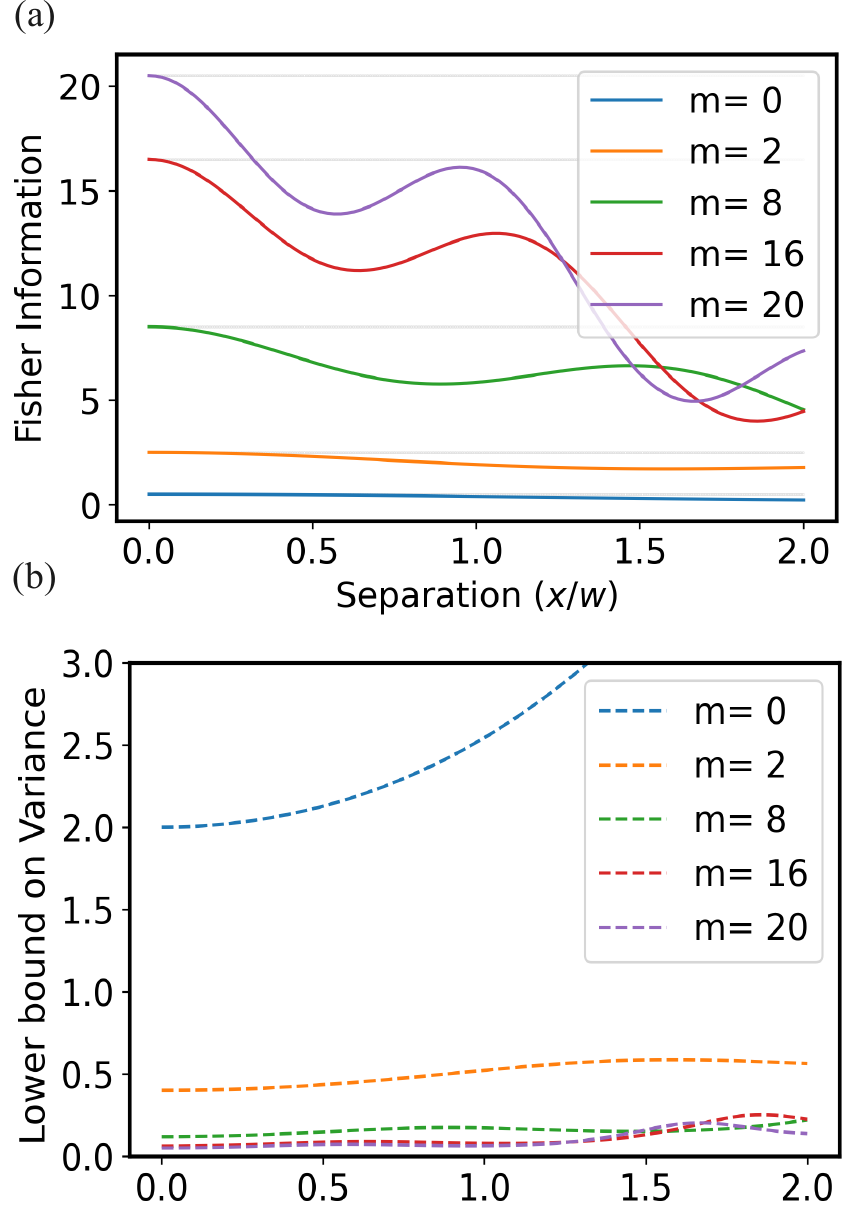}
  \caption{\textbf{Total Fisher Information and Cramér–Rao Bounds.}  
  (a) Total Fisher information obtained from projections onto modes $m-1$, $m$, and $m+1$, for an input field prepared in the $m$-th Hermite-Gauss mode. The thin gray line represents the total Fisher information obtainable from projections onto all modes for the same input, which asymptotically approaches $m + 1/2$, as discussed in the main text. (b) The corresponding Cramér–Rao bound for displacement estimates based on measurements in modes $m-1$, $m$, and $m+1$.}
  \label{VariancePlot}
\end{figure}
In what follows, we formulate the problem in more precise terms. Suppose that we send an unpolarized incoherent Hermite-Gauss beam of order $m$ through a birefringent material. The material induces a transverse displacement between the two orthogonal polarization components aligned with its fast and slow axes. 
Our goal is to estimate the transverse displacement parameter `$s$', which is proportional to the thickness `$t$' of the birefringent sample. It is well established that, under direct intensity measurements, the Fisher information vanishes as $s\to 0 $~\cite{tsang2016quantum}, a manifestation of Rayleigh’s curse in displacement sensing.

The output beams, displaced by the birefringence, emerge in the following states:
\begin{eqnarray}
   \ket{\Psi^+} =  \ket{\Psi(x+s/2)} = \ket{\mathcal{HG}_m(x+s/2)}, \\
     \ket{\Psi^-} =\ket{\Psi(x-s/2)} = \ket{\mathcal{HG}_m(x-s/2)}, \nonumber
\end{eqnarray}
where $\braket{x}{\mathcal{HG}_m(x)} = \frac{1}{\sqrt{2^m m! \sqrt{\pi}}} \mathcal{H}_m (\sqrt{2}x) e^{-\frac{x^2}{2}} $, is the $m$-th order Hermite-Gauss mode with an unit waist.
The net separation between the two states is the parameter to estimate `$s$'. For our calculations, we assume that the parameter `$s$' is normalized with respect to the waist.

The overall state $\hat{\rho}_s$, which is an incoherent mixture of a positively and a negatively displaced state about the origin, can be written as
\begin{equation} \label{Eqn:densitymatrix}
\hat{\rho}_s = \frac{1}{2}\left[\ket{\Psi^{+}}\!\bra{\Psi^{+}}+\ket{\Psi^{-}}\!\bra{\Psi^{-}}\right].
\end{equation}

For small separations, one can expand the displaced wavefunctions $\Psi^{+}$ and $\Psi^{-}$ in terms of a Taylor series with higher-order derivatives as
\begin{eqnarray} \label{eq:taylor}
\braket{x}{\Psi(x\pm s/2)} &\approx& \braket{x}{\Psi} \pm \frac{s}{2}\braket{x}{\Psi}^{\prime} + O(s^2) \\ \nonumber
&\approx& \braket{x}{\mathcal{HG}_m} \pm \frac{s}{2}\braket{x}{\mathcal{HG}_m(x)}^{\prime}.
\end{eqnarray}
Here, ${}^{\prime}$ denotes derivative with respect to $x$, i.e., $\partial/\partial x$, and all terms beyond the first-order in s are neglected. The resulting expression can be further simplified using the recurrence relation for Hermite–Gauss modes:
\begin{eqnarray} \label{eq:hermitederivativeSM}
\frac{\partial}{\partial x} \ket{\mathcal{HG}_m}=\sqrt{\frac{m}{2}}\ket{\mathcal{HG}_{m-1}}-\sqrt{\frac{m+1}{2}}\ket{\mathcal{HG}_{m+1}}. \nonumber
\end{eqnarray}
Both $\Psi^{\pm}$ are related to the adjacent Hermite-Gauss modes of $m-1$ and $m+1$. Hence if we perform projective measurement onto the modes,  $\ket{m+1} = \ket{\mathcal{HG}_{m+1}(x)}$, and $\ket{m-1}=\ket{\mathcal{HG}_{m-1}(x)}$ for the state in Equation~\eqref{Eqn:densitymatrix} the detection probabilities are given by,
\begin{eqnarray}
    P_{m \pm 1} &=& \frac{1}{2}\left|\braket{m \pm 1}{\Psi^+}\right|^2 + \frac{1}{2} \left|\braket{m \pm 1} {\Psi^-}\right|^2. \nonumber
\end{eqnarray}
Using the orthonormality of Hermite-Gauss expansion and the explicit form for the derivative, one can show that,
\begin{equation}
       P_{m+1} =  \frac{s^2}{8} (m+1), \quad \text{and} \quad 
       P_{m-1} =  \frac{s^2}{8} m .
\end{equation}
The total Fisher information per photon in this set of measurements is given by,
\begin{eqnarray}\label{FIm}
\mathcal{FI} &=& \frac{1}{P_{m+1}}\left( \frac{\partial P_{m+1}}{\partial s}\right)^2 + \frac{1}{P_{m-1}}\left( \frac{\partial P_{m-1}}{\partial s}\right)^2 \nonumber \\
&=& \frac{m+1}{2} + \frac{m}{2} \nonumber \\
&=& m + \frac{1}{2}.
\end{eqnarray}
We observe that the Fisher information is nonzero and scales linearly with the mode number $m$, with the enhancement achieved through only two measurements. This scaling in Fisher information, and consequently in sensitivity, is significantly more accessible in practice compared to schemes based on multi-photon N00N states, which become experimentally challenging to implement for photon numbers $N>2$.

The corresponding standard error in the estimates scales as,
\begin{equation}
    (\Delta \sigma)^2 \geq \frac{1}{\mathcal{FI}} \geq\frac{1}{m+1/2}.
\end{equation}
A further extension of the analysis to arbitrary separations, along with the derivation of a general expression for the Fisher information associated with the two-mode projections, is provided in Appendix A. The variances associated with different measurement schemes using various Hermite–Gauss modes, along with the corresponding Fisher information, are shown in Figure~\ref{VariancePlot}. Moreover, we consider the case where the induced displacements are coherent, with a known degree of coherence. We observe a similar scaling of Fisher information with the mode number, and note that the maximum Fisher information approaches that of the incoherent case when the two displacements are out of phase -- See Appendix B for details of the calculations.

An important consideration is that the spatial extent of a Hermite-Gauss mode of order $m$ scales as $\sim w_0 \sqrt{m + 1/2}$, where $w_0$ is the beam waist of the fundamental mode. Consequently, higher-order modes typically require a larger aperture in the detection optics. When accounting for this scaling in practical implementations, the Fisher information scales as $\mathcal{FI} \sim \sqrt{m + 1/2}$. However, in the \emph{ideal case}--where the detection aperture does not impose any constraints--the Fisher information retains its optimal scaling of $\mathcal{FI} \sim (m + 1/2)$ (see Eq.~\eqref{FIm}).
\begin{figure}[t!]
  \centering
  \includegraphics[width=\columnwidth]{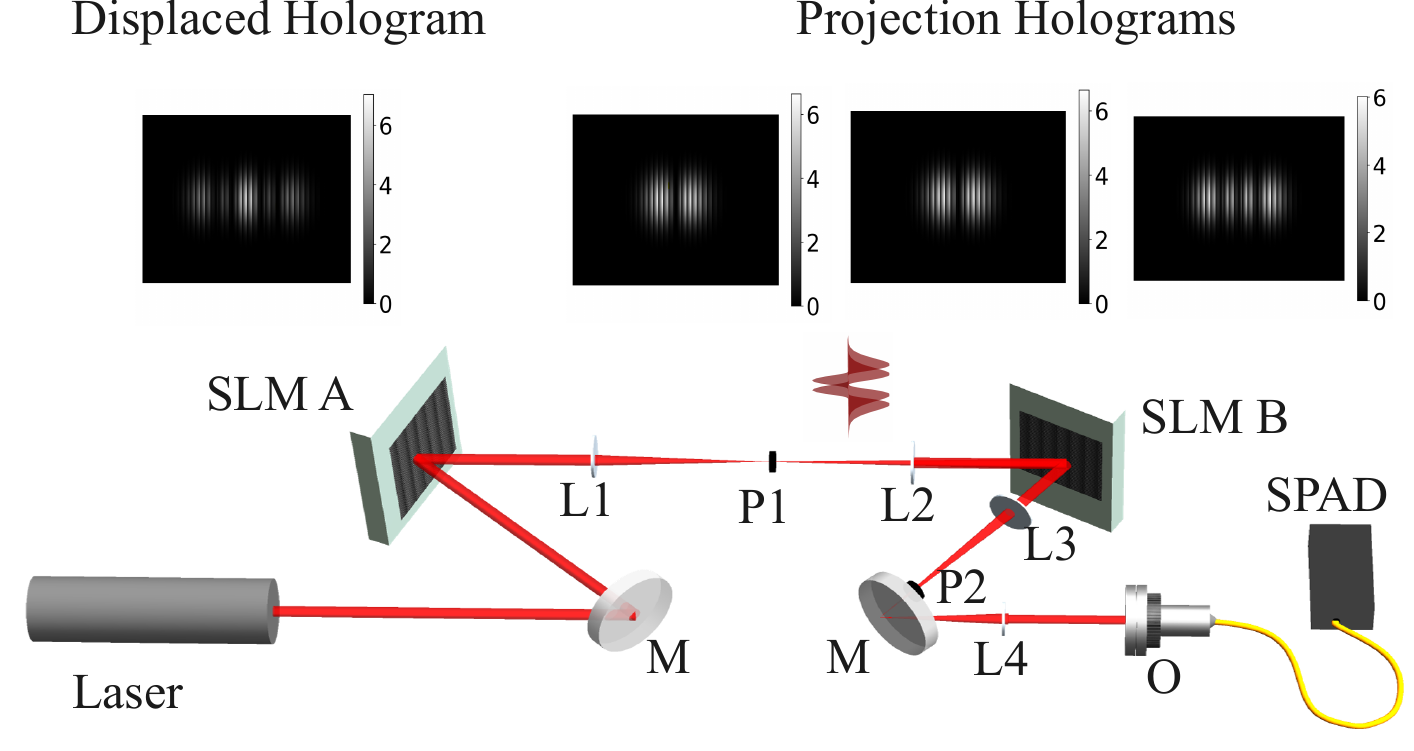}
  \caption{\textbf{Schematic of the Experimental Setup.}  
  A suitably attenuated Gaussian beam is directed onto Spatial Light Modulator A (SLM A), which generates a superposition of two displaced Hermite-Gauss modes of the desired order by displaying a custom-designed hologram. A second modulator, SLM B, together with a demagnification imaging system, performs the required mode projections.  
  Optical components are labeled as follows: M - Mirror; L1--L4 - Lenses; P1, P2 - Pinholes used to isolate the first diffraction order; O - Objective lens and translation stage; SPAD - Single-Photon Avalanche Diode.}
  \label{fig:Setup}
\end{figure}
We experimentally implemented the scheme using Hermite-Gauss modes with orders $m = 0, 1, 2, 8$, and $17$. The experimental setup is illustrated in Fig.~\ref{fig:Setup}. A suitably attenuated $808\,$nm laser beam was directed onto a Spatial Light Modulator (SLM) programmed with a custom-designed hologram~\cite{Ebrahim} to generate superpositions of displaced Hermite-Gauss modes of the desired order. The mode generated at any given instant takes the form:
\begin{eqnarray}
   \ket{\Psi} &=&  \ket{\Psi(x+s/2)} + e^{i\phi} \ket{\Psi(x-s/2)} \nonumber\\ 
   &=& \ket{\mathcal{HG}_m(x+s/2)} +  e^{i\phi} \ket{\mathcal{HG}_m(x-s/2)},
\end{eqnarray}
where $e^{i\phi}$ is a randomly chosen phase factor between the two modes. For each set of measurements corresponding to a particular mode $m$, we randomly sample the relative phase $ \phi \in [-\pi, +\pi]$ in order to simulate an incoherent mixture of the displaced modes, as described by the density matrix in Eq.~\eqref{Eqn:densitymatrix}. The resulting beam is then directed to a second Spatial Light Modulator, which displays a hologram corresponding to the appropriate projection mode. The output is coupled into a single-mode fibre and detected using a Single-Photon Avalanche Diode (SPAD) connected to a photon-counting module.
For each mode $m$, we first optimize the coupling efficiency to minimize crosstalk between adjacent modes $m-1$, $m$, and $m+1$. Displacements are then introduced in the range $s \in [0, 1.4w]$ in steps of $w/100$, where $w$ is the beam waist--this step size is close to the resolution limit of the SLM. For each displacement value, photon counts are accumulated for projections onto modes $m-1$, $m$, and $m+1$, across randomly sampled phase values.
From the photon count statistics, we extract mode probabilities and estimate the displacement using a Maximum Likelihood Estimation (MLE) approach. The variances in the estimated displacements are then computed and plotted in Fig.~\ref{ExpPlots}. We observe an order-of-magnitude improvement in estimation precision when comparing lower-order modes with higher-order cases, particularly for $m = 8$ and $m = 17$.\newline
In summary, we have demonstrated that probing incoherent transverse displacements using higher-order Hermite-Gauss (HG) modes leads to a sensitivity enhancement that scales linearly with the mode order, provided that projective measurements are performed onto neighboring spatial modes. This stands in contrast to conventional direct imaging techniques, which suffer from diminished Fisher information in the small-displacement regime.
\begin{figure}[t!]
  \centering
  \hspace*{0 cm}
  \includegraphics[width=\columnwidth]{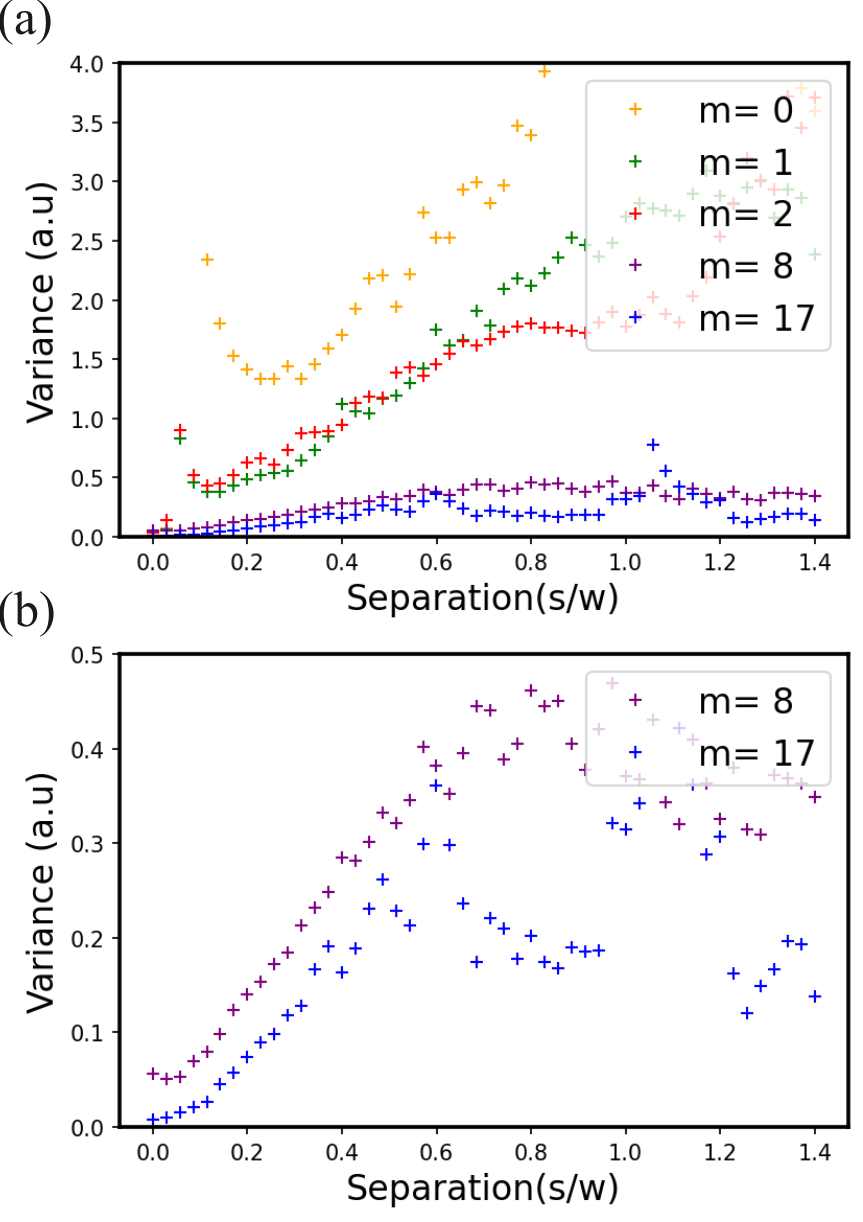}
  \caption{\textbf{Experimentally Measured Variances.}  (a) Variances in displacement estimates extracted from optimal projections for input modes $m=0$, $1$, $8$, and $17$.  (b) A magnified view of the variances for $m=8$ and $m=17$, highlighting the improvement in precision relative to the fundamental mode $m=0$ shown in (a). An order-of-magnitude reduction in estimation error is observed for the higher-order modes.}
  \label{ExpPlots}
\end{figure}
We presented a proof-of-principle experimental realization, observing an order-of-magnitude improvement in sensitivity compared to standard SPADE measurements when using HG modes with order $m \geq 8$. Our findings also offer a reinterpretation of the recent experiment reported in Ref.~\cite{grenapin2023superresolution}, where SPADE was applied to a correlated photon-pair source, and the transverse displacement was introduced on the idler photon. In that work, the observed Fisher information scaled with the square root of the Schmidt number, providing an effective estimate for the number of entangled HG modes comprising the biphoton state. Under the Klyshko picture, detection in mode $m$ on the signal arm corresponds to post-selecting a process in which the idler photon is prepared in the same HG mode, undergoes a transverse displacement, and is subsequently projected onto neighboring modes -- an operationally equivalent scenario to our present experiment. Future investigations may explore this approach with pump beams prepared in higher-order HG modes, potentially revealing a pathway to quantum-enhanced displacement sensing.
In our implementation, we employed sequential projective techniques based on amplitude flattening, which inherently involve a trade-off between mode-dependent losses and inter-mode crosstalk~\cite{bouchard2018measuring}. However, advanced mode projection strategies, such as those utilizing multi-plane light converters, have recently been demonstrated for SPADE at the single-photon level~\cite{rouviere2024ultra,santamaria2024single}, and could serve as a robust alternative for practical deployment. We therefore believe that the techniques demonstrated in this work, combined with state-of-the-art optical technologies, hold strong promise for the development of compact and precise devices capable of detecting minute displacements--particularly in applications involving nanoscale systems~\cite{choi2024quantum} and high-resolution imaging of birefringent materials.

\emph{Acknowledgments--} This work was supported by the Canada Research Chair (CRC) Program, the Joint Center for Extreme Quantum Photonics (JCEP) of NRC-uOttawa through the Quantum Sensors Challenge Program at the National Research Council of Canada, and the Quantum Enhanced Sensing and Imaging Alliance Consortium Quantum Grant (QuEnSI).

\bibliography{Higherordersuperres}

\newpage
\appendix

\section{Fisher Information at Arbitrary Separation}
Here, we extend the Fisher Information Analysis to the regime of arbitrary large displacements and show that the sensitivity is enhanced even at large separations depending upon the mode order $m$. At large separations, the mth mode in addition to $m-1$, and $m+1$ also contains a significant amount of Fisher information for the separation parameter. Thus our optimal projections for large enough separations is
\begin{eqnarray}
    \ket{m+1} &=& \ket{\mathcal{HG}_{m+1}(x)} \\
    \ket{m}  &=& \ket{\mathcal{HG}_m(x)} \\
     \ket{m-1} &=& \ket{\mathcal{HG}_{m-1}(x)}
\end{eqnarray}
We first evaluate the probability for projection unto $\ket{m+1}$
\begin{eqnarray} \label{eq:P}
    P_{m + 1} &=& \frac{1}{2}\left|\braket{m + 1}{\Psi^+}\right|^2 + \frac{1}{2} \left|\braket{m + 1} {\Psi^-}\right|^2 \nonumber \\ \nonumber
     &=& \frac{1}{2}\left|\braket{m + 1}{m^+}\right|^2 + \frac{1}{2} \left|\braket{m + 1} {m^-}\right|^2\\
\end{eqnarray}
where,
$\ket{m^\pm}$ refer to the displaced HG mode, $\mathcal{HG}_m(x\pm s/2)$. 
We proved in our previous work \cite{grenapin2023superresolution} that for $m\geq n$
\begin{equation} \label{eq:analyticalSM}
\braket{m}{n^\pm} = \sqrt{\frac{n!}{m!}}2^{\frac{n-m}{2}}(\mp s)^{m-n}e^{-\frac{s^2}{4}}\mathcal{L}_n^{m-n}(\frac{s^2}{2})
\end{equation}
which gives us,
\begin{equation} \label{eq:analyticalSM}
\braket{m+1}{m^\pm} = \sqrt{\frac{1}{m+1}}2^{\frac{-1}{2}}(\mp s)^{1}e^{-\frac{s^2}{4}}\mathcal{L}_m^{1}(\frac{s^2}{2}).
\end{equation}
Substituting this in Eqn \eqref{eq:P} and simplifying we get,
\begin{eqnarray}
    P_{m+1} = \frac{1}{8(m+1)} e^{-\frac{s^2}{8}}s^2\left(\mathcal{L}_m^{1}\left(\frac{s^2}{8}\right)\right)^2.
\end{eqnarray}

And similarly, the probability for projection unto $\ket{m-1}$, and $\ket{m}$ can be evaluated and the final expression assumes the form,
\begin{eqnarray}
    P_{m-1} &=& \frac{1}{8m} e^{-\frac{s^2}{8}}s^2\left(\mathcal{L}_{m-1}^{1}\left(\frac{s^2}{8}\right)\right)^2 \\
    P_{m} &=& e^{-\frac{s^2}{8}}\left(\mathcal{L}^0_m\left(\frac{s^2}{8}\right)\right)^2.
\end{eqnarray}
The Fisher information in these two projections then can be evaluated as,
\begin{eqnarray}
\mathcal{FI} &=& \sum_{i=1,0,-1}\frac{1}{P_{m+i}}\left( \frac{\partial P_{m+i}}{\partial s}\right)^2 \nonumber \\
&=& \frac{e^{-\frac{s^2}{8}} \left( s^2 \mathcal{L}_{m-1}^2\left(\frac{s^2}{8}\right)+\left(0.5 s^2-4\right) \mathcal{L}_m^1\left(\frac{s^2}{8}\right)\right){}^2}{32(n+1)} \nonumber \\
&+& \frac{e^{-\frac{s^2}{8}}  \left(s^2 \mathcal{L}_{m-2}^2\left(\frac{s^2}{8}\right)+\left(0.5 s^2-4\right) \mathcal{L}_{m-1}^1\left(\frac{s^2}{8}\right)\right){}^2}{32n} \nonumber \\
&+& \frac{e^{-\frac{s^2}{8}} \left(s \mathcal{L}_m\left(\frac{s^2}{8}\right)+2s \mathcal{L}_{m-1}^1\left(\frac{s^2}{8}\right)\right){}^2}{16}. \nonumber
\end{eqnarray}
The Fisher Information approaches the previously derived value of $m+ \frac{1}{2}$ for $s \to 0$. The Cramer-Rao bounds for the variances can be derived using
$
    (\Delta \sigma)^2 \geq \frac{1}{\mathcal{FI}}.
$
The variances for different measurements using different modes along with the Fisher Information are plotted in Figure \ref{VariancePlot}.
\newline

\section{Fisher Information for coherent separations}
Here we extend the calculation of fisher information for coherent displacements. Say the material introduces coherent displacement with a coherence parameter $\gamma$, where 
$\gamma = e^{i\phi}$, between the positively and negatively displaced states. The normalized state after such interaction becomes,
\begin{eqnarray}
   \ket{\Psi} &=& \frac{1}{\sqrt{2}} \left( \ket{\Psi^+} + \gamma \ket{\Psi^-} \right)\nonumber \\
   &=& \frac{1}{\sqrt{2}}\left( \ket{\mathcal{HG}_m(x-s/2)} + \gamma \ket{\mathcal{HG}_m(x+s/2)} \right).\nonumber
   \label{coherentstate}
\end{eqnarray}

For small separations, one can expand the displaced wavefunctions in terms of a Taylor series with higher order derivatives as before,
\begin{eqnarray} \label{eq:taylor}
\braket{x}{\Psi(x\pm s/2)} &\approx& \braket{x}{\Psi} \pm s/2 \braket{x}{\Psi}^{\prime} + O(s^2) \\
&\approx& \braket{x}{\mathcal{HG}_m} \pm s/2 \braket{x}{\mathcal{HG}_m(x)}^{\prime}
\end{eqnarray}

Hence if we perform projective measurement onto the modes,
\begin{eqnarray}
    \ket{m+1} = \ket{\mathcal{HG}_{m+1}(x)}, \\
     \ket{m-1} = \ket{\mathcal{HG}_{m-1}(x)},
\end{eqnarray}
 for the state  in Equation ~\ref{coherentstate} the detection probabilities are given by,
\begin{eqnarray}
    P_{m \pm 1} &=& \frac{1}{2}\left|\braket{m \pm 1}{\Psi^+} + \gamma \braket{m \pm 1} {\Psi^-}\right|^2 \nonumber
\end{eqnarray}
\begin{figure}[t!]
  \centering
  \hspace*{0 cm}
  \includegraphics[width=\columnwidth]{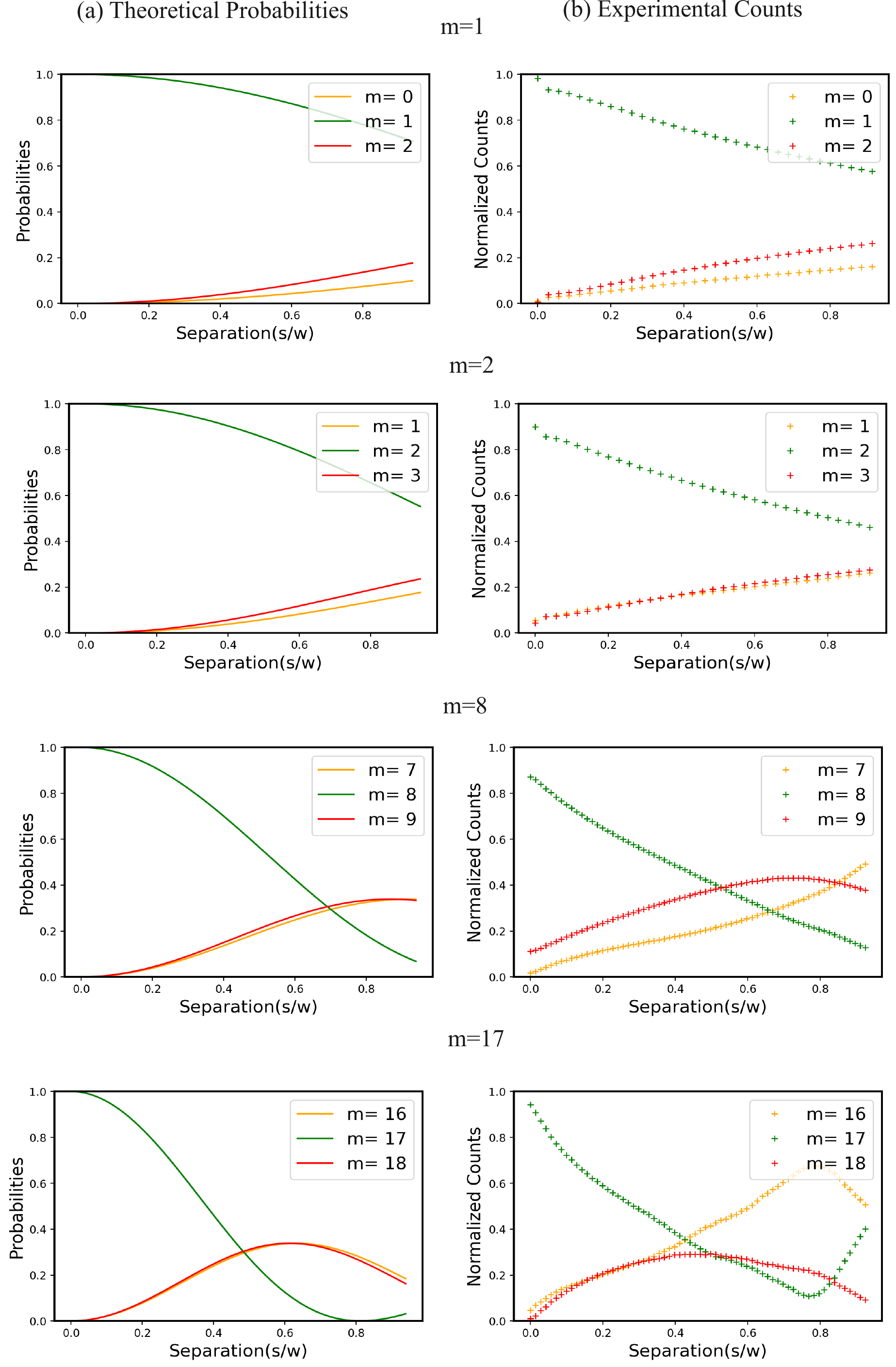}
  \caption{Theoretical Probabilities (a) versus the Experimental counts (b) for the $m-1$, $m$, and $m+1$th modes for $m=1$, $m=2$, $m=8$, and $m=17$. }
  \label{ExpPlots}
\end{figure}

Using the orthonormality of Hermite Gauss expansion and the explicit form for the derivative, one can show that,
\begin{equation}
       P_{m+1} =  \frac{s^2}{8} (m+1) (1-\gamma)^2, \quad \text{and} \quad 
       P_{m-1} =  \frac{s^2}{8} m (1- \gamma)^2.
\end{equation}
The total Fisher information per photon in this set of measurements  is given by,
\begin{eqnarray}
\mathcal{FI} &=& \frac{1}{P_{m+1}}\left( \frac{\partial P_{m+1}}{\partial s}\right)^2 + \frac{1}{P_{m-1}}\left( \frac{\partial P_{m-1}}{\partial s}\right)^2 \nonumber \\
&=& \left(\frac{m+1}{2} + \frac{m}{2}\right)(1-\gamma)^2 \nonumber \\
&=& \left(m + \frac{1}{2}\right)\left(1-\gamma\right)^2.
\label{FImcoherent}
\end{eqnarray}
We see that the Fisher information is maximized and approaches the incoherent case as $\gamma = -1$ or $\phi = \pi$, i.e. when the two displacement are out of phase with each other. Whereas in case of displacements that are in phase, the Fisher Information is zero. We see that the enhancement obtained from the mode order $m$ is still present.

\end{document}